\documentclass[4paper,12pt]{article}
\usepackage[brazilian,british]{babel}
\usepackage[utf8]{inputenc}
\usepackage[T1]{fontenc}
\usepackage{multicol, blindtext, graphicx}
\usepackage{graphicx}
\usepackage{epsfig}
\usepackage[usenames]{color}
\usepackage{amssymb}
\usepackage{amsmath}
\usepackage{cite}
\usepackage{times}
\usepackage{authblk}
\usepackage{setspace}
\usepackage{appendix}
\usepackage{lineno}
\usepackage{multicol}
\usepackage[left=2cm,top=3cm,right=1.5cm,left=1.5cm,foot=2.5cm]{geometry}
\usepackage[runin]{abstract}
\usepackage[compatibility=false,labelfont=it,textfont={bf,it}]{caption}
\abslabeldelim{:\,} 

\addto\captionsbritish{
}

\title{A proposal for an active methodology for physics labs\\
}
\author{Suzane F. Pinto}
\affil{\small\textit{Programa de graduação em Engenharia de Computação,
Universidade Federal de Ouro Preto, 35931-008, Brasil}}
\author[2]{Ronan S. Ferreira\thanks{ronan.ferreira@ufop.edu.br}}
\affil[2]{\small\textit{Departamento de Ciências Exatas e Aplicadas,
Universidade Federal de Ouro Preto, 35931-008,  Brasil}}
\author[2]{Miguel M.Costa}
\author{Agmael M. Silva}
\affil{\small\textit{Campus Prof. Antonio Giovanne Alves de Sousa, Universidade Estadual do Piauí, 64260-00, Brasil}}
\date{}
\newenvironment{multiabstract}[1]
  {\renewcommand{\abstractname}{#1}\begin{abstract}}
  {\end{abstract}}
\begin{document}
\maketitle
    \maketitle
\selectlanguage{british}
    \begin{multiabstract}{Abstract}
    Active methodologies aim to develop a critical sense of what is learned, relating theoretical concepts to the practical environment. In this work, we propose an active teaching-learning methodology for laboratory classes in which the student has the autonomy to propose scripts and equipment, instead of following a practice roadmap already defined (built by the teacher or made available by manufacturers for their science kits), in accordance with the theoretical knowledge acquired. The objective is to encourage the student to be the protagonist in experimental activities, based on the theoretical knowledge acquired in the classroom. In this way, we split the method into three parts, namely: ({\em i}) Theoretical exposition, ({\em ii}) theoretical seminar and proposition of the experimental script and ({\em iii}) seminar for the exposition of the experiment carried out. Each of these steps is guided by one or more professional skills, such as: innovation, creativity, proactivity, protagonism, critical sense and scientific thinking, aiming to bring the academic environment to the professional environment.\\
\textbf{Keywords}: Physics education, Teaching methodologies, Active methodology.
\end{multiabstract} 

\begin{multicols}{2}
\section{Introduction}
\label{sec:intro}
    With countless interactive channels available on social media, proposing new methodologies in the classroom that encourage the student to change the passive role for the protagonist in his own learning is a challenge for the teaching activity \cite{otero2016100,otero2017past}, particularly on the lab environment \cite{smith2020expectations,kestin2020comparing}. An important point is that for this to happen, a two-way flow is suggested: teacher and student. Speeches commonly verbalized by both teachers and students exemplify why the change needs to occur in two senses. On the one hand, we have traditional methodologies that gain the status of routine classes, from the teaching point of view, and little involving when viewed from the students' perspective. On the other hand, it is clear that the use of new technological resources, which at first glance would be the promise for more dynamic classes, in fact do not seem to change this scenario of permanent and collective dissatisfaction. The application of technology alone does not guarantee a high standard of learning \cite{worthington2015provide}, although can be used as an ally in the learning process \cite{kestin2020comparing}.

    In addition to this challenge for teaching activity, there is another aspect that is increasingly present: the academy-business relationship. We can think of it in at least two ways. First, how has the academic environment been preparing its human resources according to what the job market has been looking for? Second, how to prepare students so that, even in the academic environment, they develop practical skills for the promotion of {\em Jr. Companies}, {\em Startups} and thus draw attention and promote partnerships between academy and companies?

    Jr Companies are non-profit companies whose main objective is to support practical learning in university education through projects, learning by management and entrepreneurial culture. Startups are companies that develop and provide services and products exploring innovative activities in the market in which they operate. In these companies, technology and innovation are present at all levels: strategic, operational and tactical. For example, the business model - strategic level - is based on an economic model that aims to reach a considerable number of customers and generate profit on a scale without a proportional increase in the costs of the operation. Within the Jr and Startups Company, we have agile development environments that are usually built by teams that have autonomy, seeking objectives and goals that each employee can achieve in his time.

    Bearing in mind that the search for professionals capable of acting with autonomy and creativity has become a paradigm, how to prepare the student still within the university? It is here that we can see the importance of practical and collective work to be developed in the laboratories.

    On the one hand, the job market has been seeking professional skills such as {\em innovation / creativity, proactivity / protagonism, critical sense / scientific thinking}, etc. On the other hand, our laboratories are full of ``science kits'': pre-assembled prototypes, substantially immune to human errors and overly detailed scripts about their execution: a kind of infallible guide in order to obtain the best results - therefore, the smallest mistake. In this way, the student remains in the passive role in his own learning, following a mere laboratory algorithm.
      
    Active learning methodologies  \cite{coll2003psicologia,diesel2017principios} seek to promote meaningful learning that requires, in the first place, a systematization of teaching that is capable of involving the student as a protagonist of their learning. In this way, such methodologies aim to develop a critical sense of what is learned, as well as skills to relate theoretical concepts to the real world \cite{holmes2015teaching,walsh2019quantifying,viennot2018activation}. It is important to think about methodologies for an educational practice that seeks the formation of an active professional, able to {\em learn to learn}.

    According to Bergamo\cite{bergamo2010uso}, traditional expository classes are very tiring for both students and teachers, and in most cases they are not accompanied with the practical part, in order to make a connection between theoretical concepts and real situations. In general, we have a class in which only the teacher acts by exposing and sometimes imposing. Therefore, he is the only protagonist and we have, in most cases, a lack of interaction by students because they do not absorb the content or even the simple lack of interest in the content exposed.

    To contribute to the proposal of active teaching methodologies for physics laboratory classes, we propose in this work a method in which the student, instead of following a practical script already defined (built by the teacher or made available by manufacturers for their science kits), has the autonomy to propose experiments (scripts and equipment), in accordance with the theoretical knowledge acquired and in the light of the scientific method.

\section{Methodology}
\label{sec:met}
    The methodology we present in this work has as a guide to encourage the student to be the protagonist of experimental activities, based on the theoretical knowledge acquired in the classroom. In this way, we split the method into three parts, namely: ({\em i}) Theoretical exposition, ({\em ii}) Proposal seminar, with the choice of theme and an experimental design proposition and ({\em iii}) Final seminar for the exposition of the experiment carried out. In this methodology, only item ({\em i}) is the responsibility of the teacher, while items ({\em ii}) and ({\em iii}) depend on the protagonism of the students (evidently, with teacher guidance). Note that this methodology differs from the commonly used laboratory methodologies, since they generally assume the following protocol: ({\em i}) theoretical exposition, ({\em ii}) exposition of the experimental script, ({\em iii}) execution of the experiment and ({\em iv}) report of the experimental activity. In this way, teachers carry out items ({\em i}) and ({\em ii}). In the next sections, we will detail items ({\em i}), ({\em ii}) and ({\em iii}) of our proposal.

\begin{figure*}[ht]
	\centering
	\includegraphics[scale=1.0]{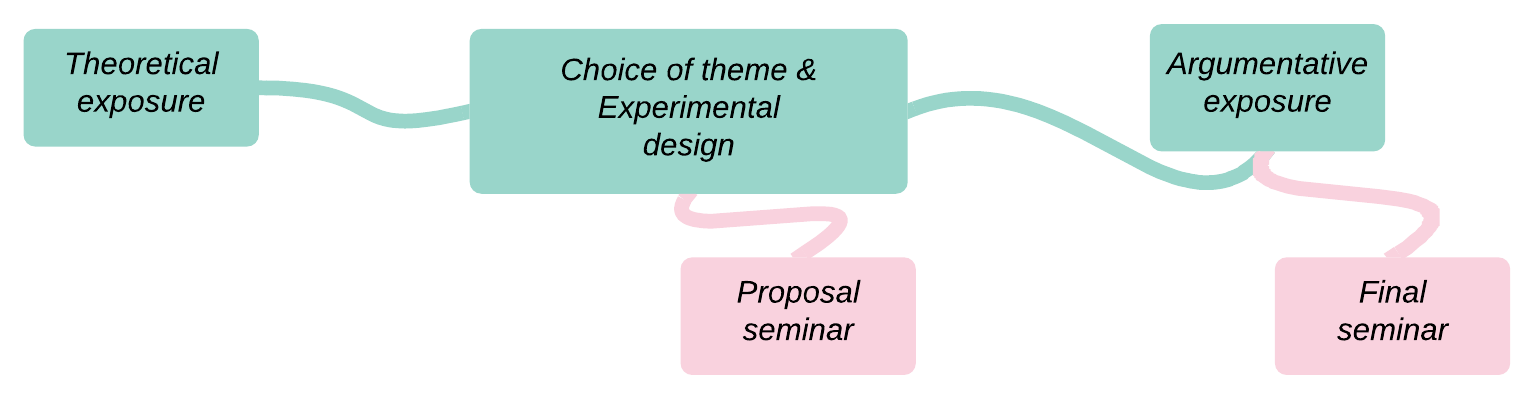}
	\caption{\label{FIG1}\textbf{Diagram of the model}. Each frame means a stage performed in the proposed methodology. It is highlighted students' seminars on activities performed.}
\end{figure*}

\subsection{Theoretical exposure}
\label{sub:1}
    It is at the stage of the theoretical exposure that the teacher will take the lead in the student's learning process. Since in the classroom environment we find a vast diversity of people, each with their own way of thinking, reasoning, interpreting, and acting, several pedagogical strategies can be used.

    As references, we can mention already consolidated methodologies such as Peer instruction (PI) or Peer learning (AEP), STEAM and Constructivist Spiral (EC). The AEP allows students to assume the roles of protagonists during classes, in moments of debates with colleagues, when they are solving activities related to the topics under study. The teacher has the role of mediating and guiding the discussions between them  \cite{de2017associaccao}. EC is based on the idea of dividing the process into stages and carrying them out in a circular manner. Steps like identifying problems, formulating explanations, elaborating questions, constructing new meanings and evaluating processes and products \cite{lima2016espiral}. STEAM is an acronym for {\em Science, Technology, Engineering, Arts and Mathematics}. It is considered an integrated and project-based methodology, which aims to encourage interdisciplinarity and always focus on the practical application of the learning developed within the classroom \cite{hardoim2019educaccao}. 

    The objective of this stage is to expose the student to the technical-theoretical knowledge necessary for him to be able to perform the activities proposed in the following steps. 

\subsection{Proposal seminar}
    This is the stage in which the student must propose an experiment to verify one or more concepts discussed in the previous stage - section \ref{sub:1}. In comparison with the methodology commonly applied, this step would be an alternative to the initial part of a laboratory activity, in which the student usually receives the script for the practical class. For our proposal, the student must present the necessary materials, as well as procedures and methodology to be used. In other words, the student must design a project himself in order to verify one or more theoretical concepts. Note that at this point there is an important exchange: The old “script” takes on a more professional role with the concept of developing a “project”. At this point, it is also worth mentioning the link between our proposal and the STEAM methodology. The student may be encouraged to use knowledge already acquired in other disciplines. In effect, the teacher assumes the role of tutor here, encouraging and pointing out the relationship of his discipline with others in the curriculum of the student's qualification.

    In this stage, aspects such as creativity and critical sense will be in focus, since students must propose materials / equipment to achieve an outlined objective. Students will be able to propose from the traditional equipment of a physics laboratory (scale, measuring tape, objects with different masses, calipers, etc.) and even alternative materials / equipment. As an example, we can mention smartphones, digital cameras, recycled materials, toys etc. Thus, the place to perform the proposed experiment is also flexible. It is worth remembering that the proposal presented by the students in this seminar will give the teacher conditions to evaluate possible risks regarding the realization of the experiment. 
    
    Another interesting point is that the physics laboratory is available to students. They can / should be encouraged to also discuss with the professional laboratory technician about their experimental proposals. Again, the role of the teacher at this stage as a tutor is highlighted, assisting in activities, solving doubts, assessing the feasibility of the proposed projects and leaving the protagonism of the activity with the student.

\subsection{Final seminar}
    At this stage the student is in full prominence, since it is the stage in which he will defend his experimental project, proposed in the previous stage.

    If compared to a common practice report, this step would be what we call ``results and discussions'', however, in a dynamic way. Both the results and the discussions will be thought and commented on in a seminar format, thus opening space for constructive criticisms about the work developed. Evidently, the teacher assumes the role of mediator, encouraging the participation and discussion of all listeners in the class. Aspects such as protagonism, improved orality and the search for proactivity are intensified.

    This gives students the opportunity to experience an environment in the university in which they will be inserted in the job market in the future. That is why it is of great importance to encourage them to be protagonists, so that they are prepared to deal with environments where the least important thing is to follow orders but to perform tasks in a timely manner, with autonomy and responsibility.

\section{Results}
    To quantitatively estimate the reception and evaluation of students regarding the methodology proposed in section \ref{sec:met}, we developed a survey (discussed below) based on a Likert scale fashion. This verification scale consists of assuming a construct and developing a set of statements relevant to its definition, for which the interviewees will state their degree of agreement. Essentially, this is a one to five point scale capable of inferring more information than using competing methods. It can be defined as a type of ``attitude scale'', in which the degree of agreement is verified in relation to a given questioning \cite{bermudes2016tipos,appolinario2007dicionario}. The typical format of responses, from 1 to 5, accessible to the interviewee is: 1. {\em Totally disagree}; 2. {\em Partially disagree}; 3. {\em Indifferent}; 4. {\em Partially agree}; 5. {\em Totally agree}. Appendix A shows the poll used. 

    The purpose of the statements used in the survey was to address perception for those skills mentioned in section \ref{sec:intro}, namely: protagonism, creativity, critical sense and responsibility. In addition, evaluate a measure of success for the use of the proposed methodology. 

\begin{figure*}[ht]
	\centering
		\includegraphics[scale=0.4]{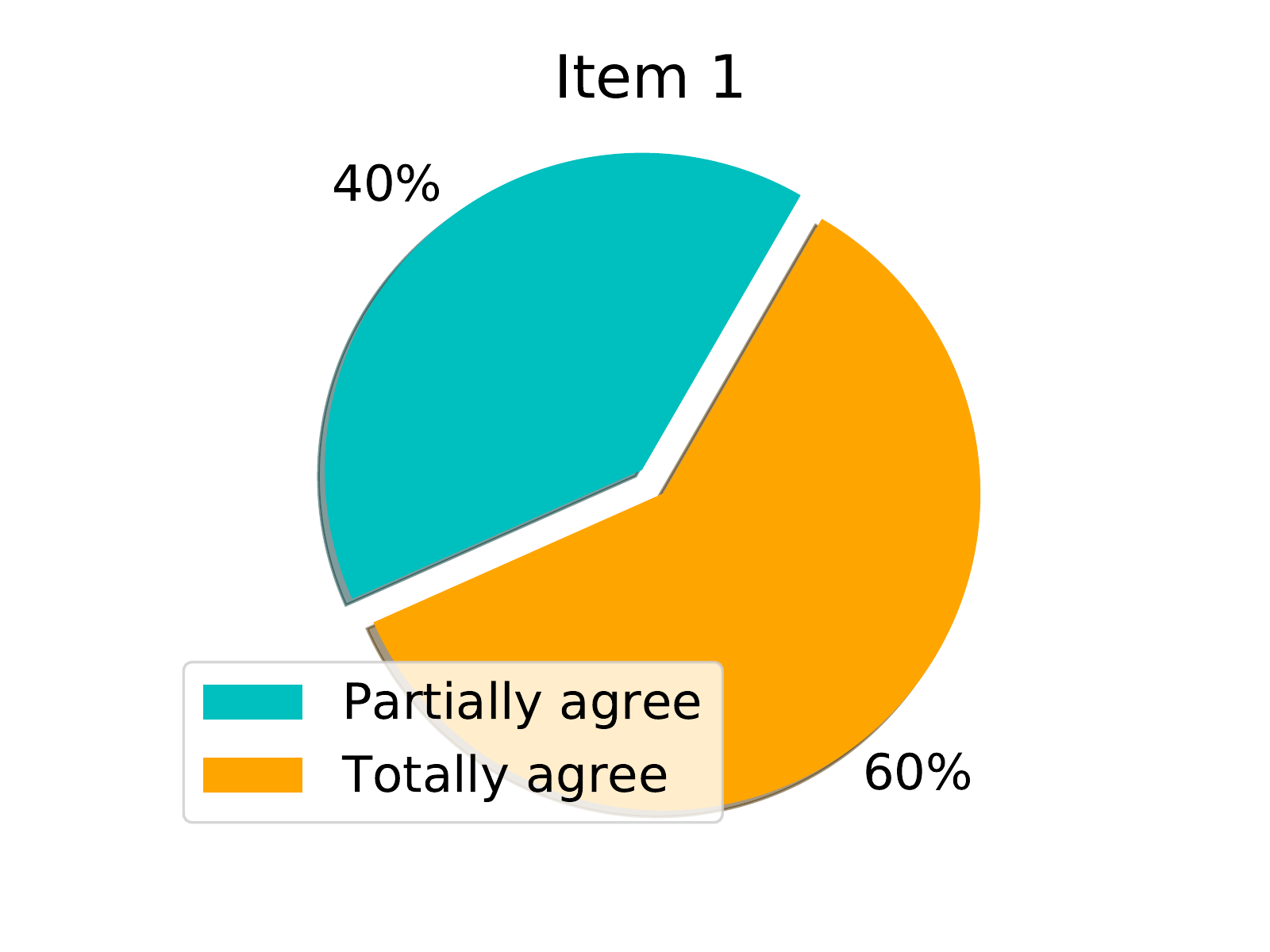}\includegraphics[scale=0.4]{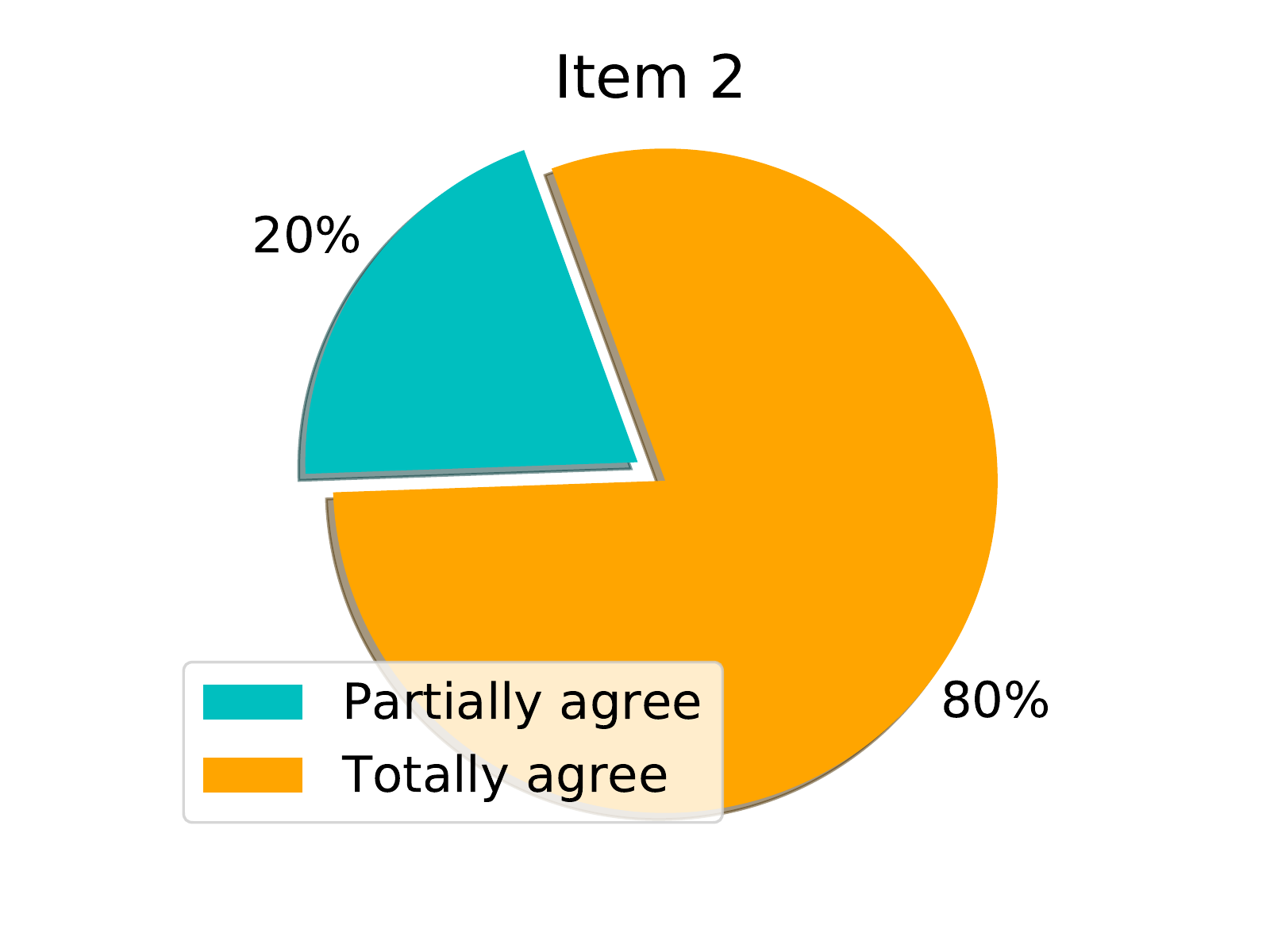}\includegraphics[scale=0.4]{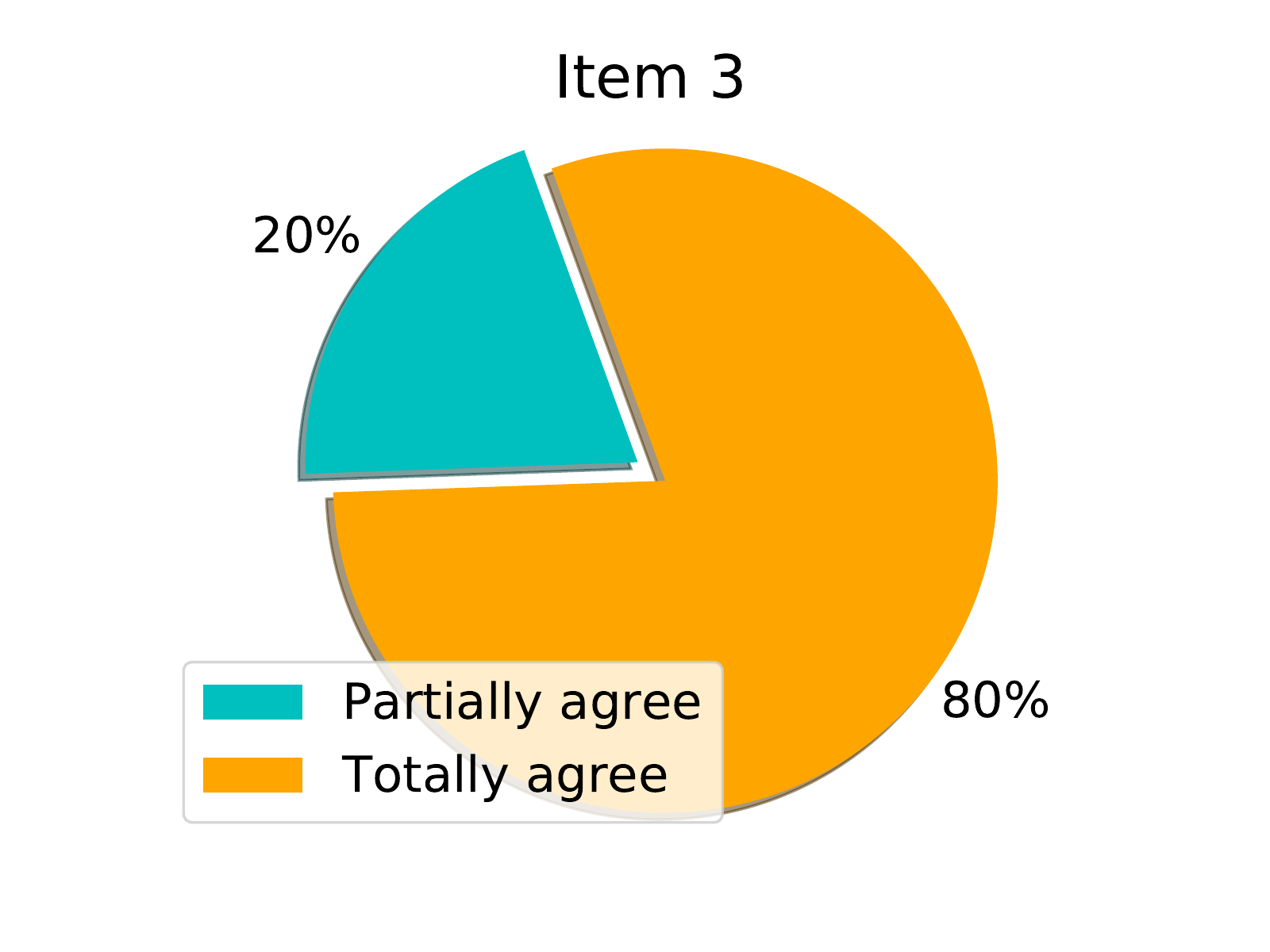}\\
		\includegraphics[scale=0.4]{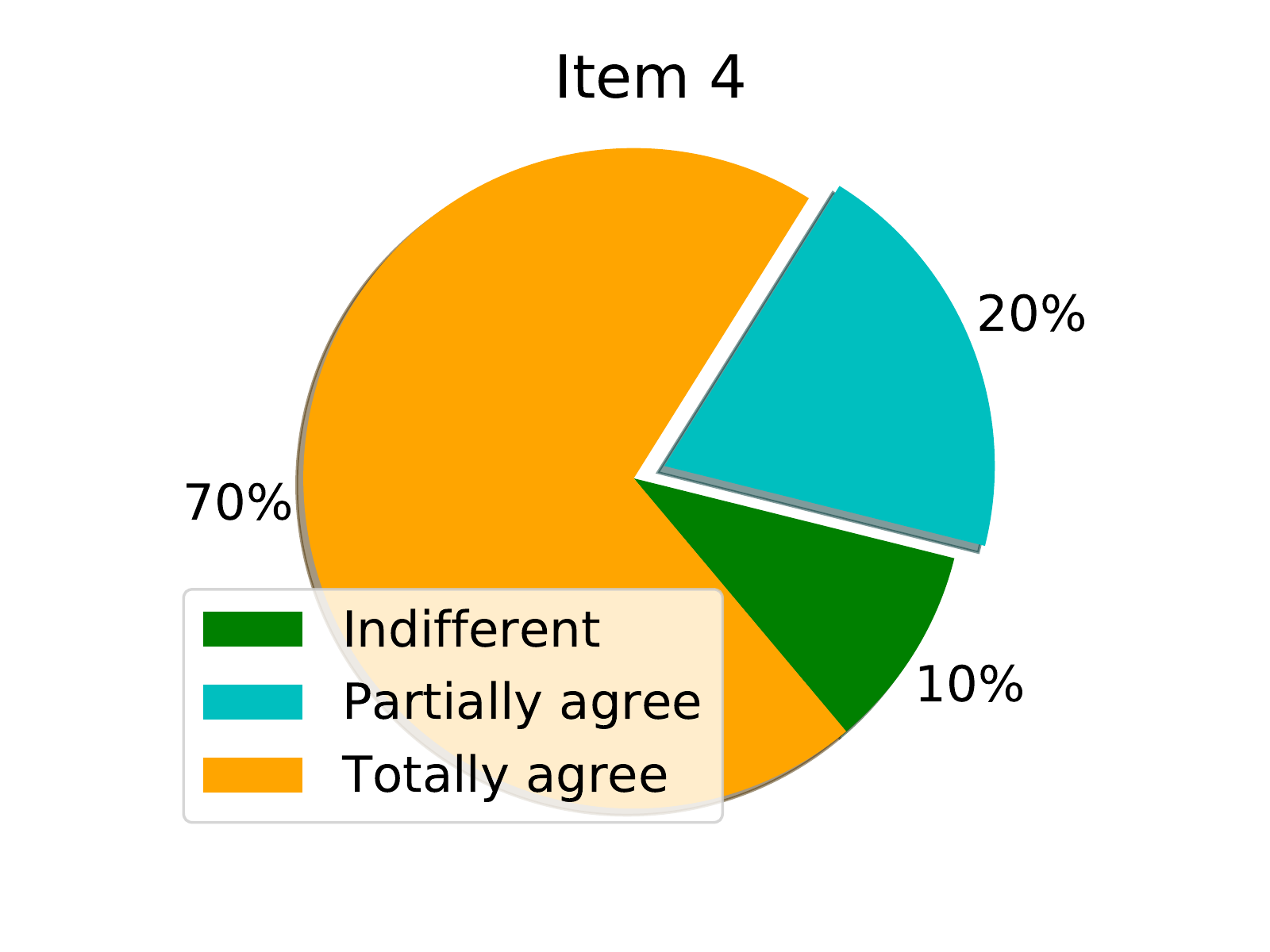}\includegraphics[scale=0.4]{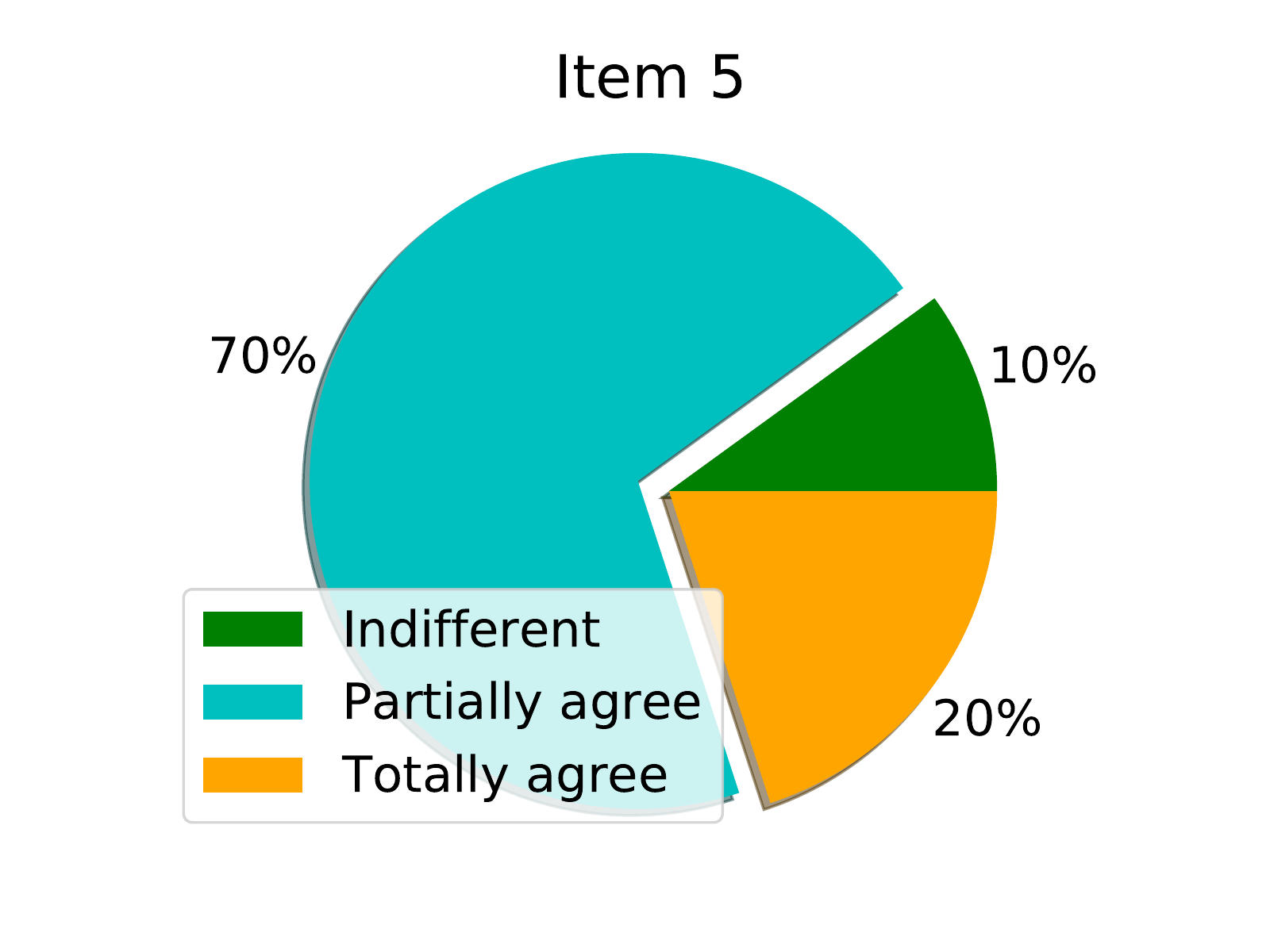}\includegraphics[scale=0.4]{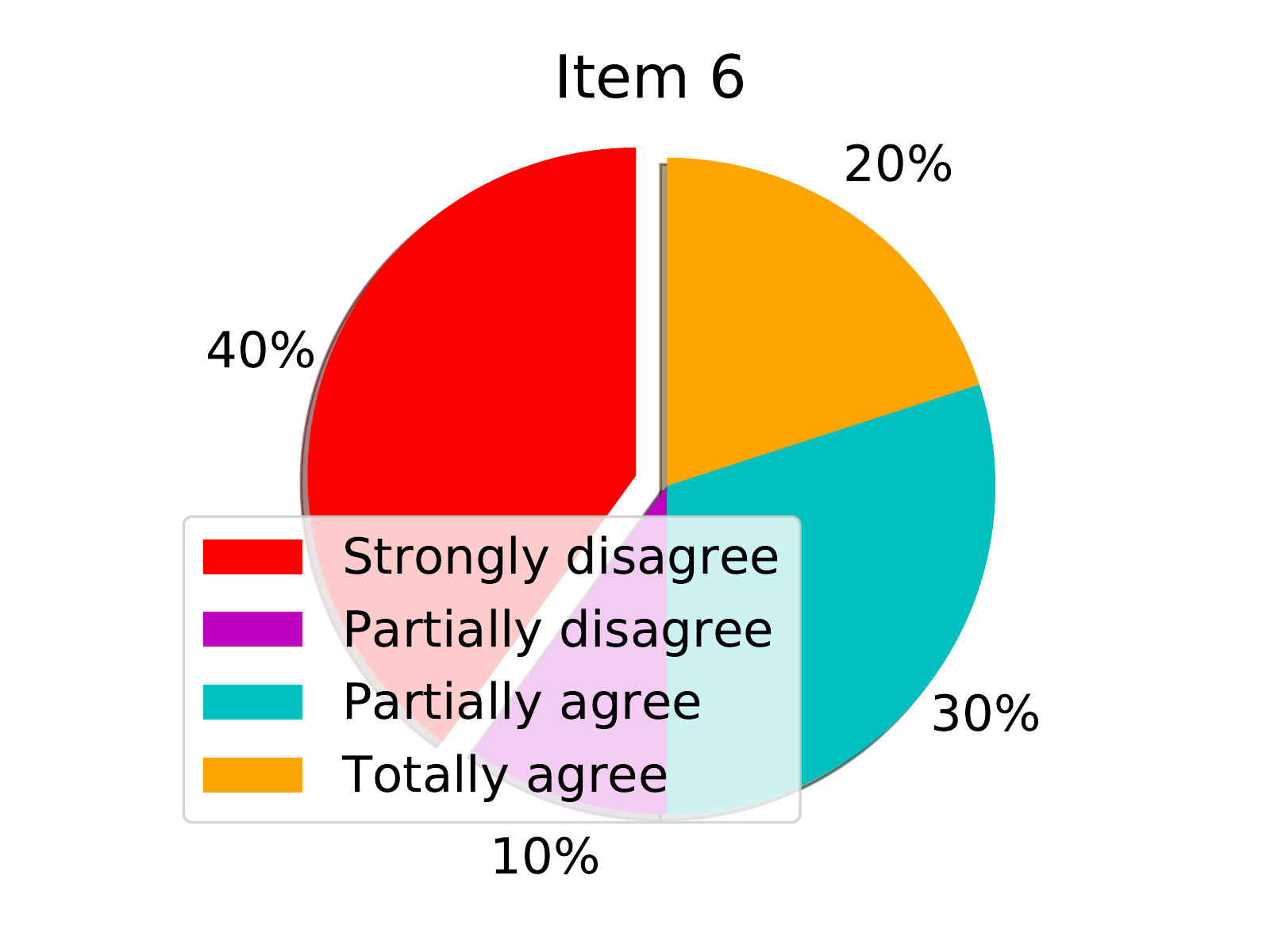}
	\caption{\label{fig2} \textbf{1\textsuperscript{st} Block of results}. The numbering of each of the graphs in this figure corresponds to the numbering of each question, from 1 to 6, of the applied questionnaire - see Appendix~\ref{Anx:A}.} 
\end{figure*}

\begin{figure*}[ht]
	\centering
	\includegraphics[scale=0.4]{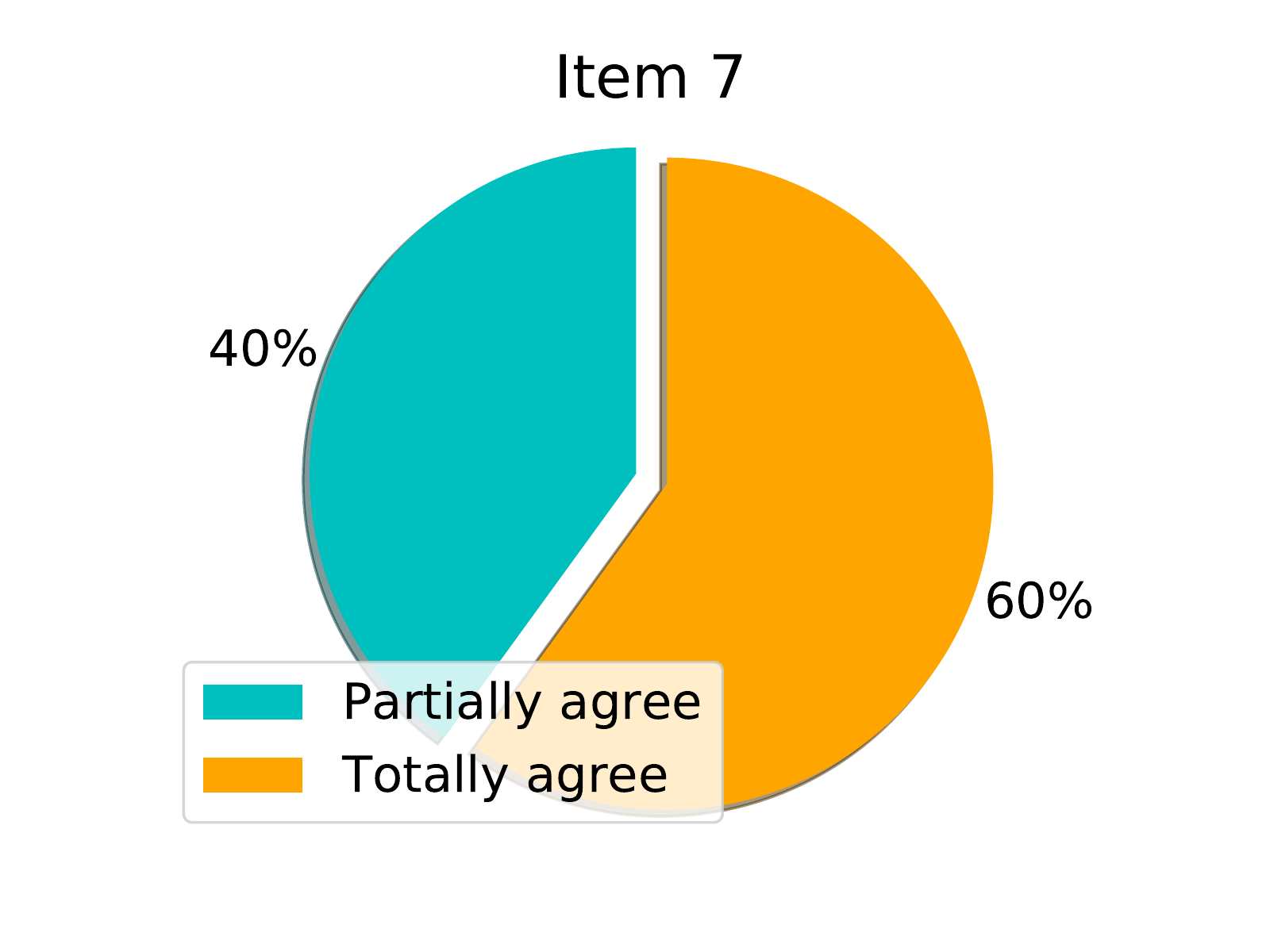}\includegraphics[scale=0.4]{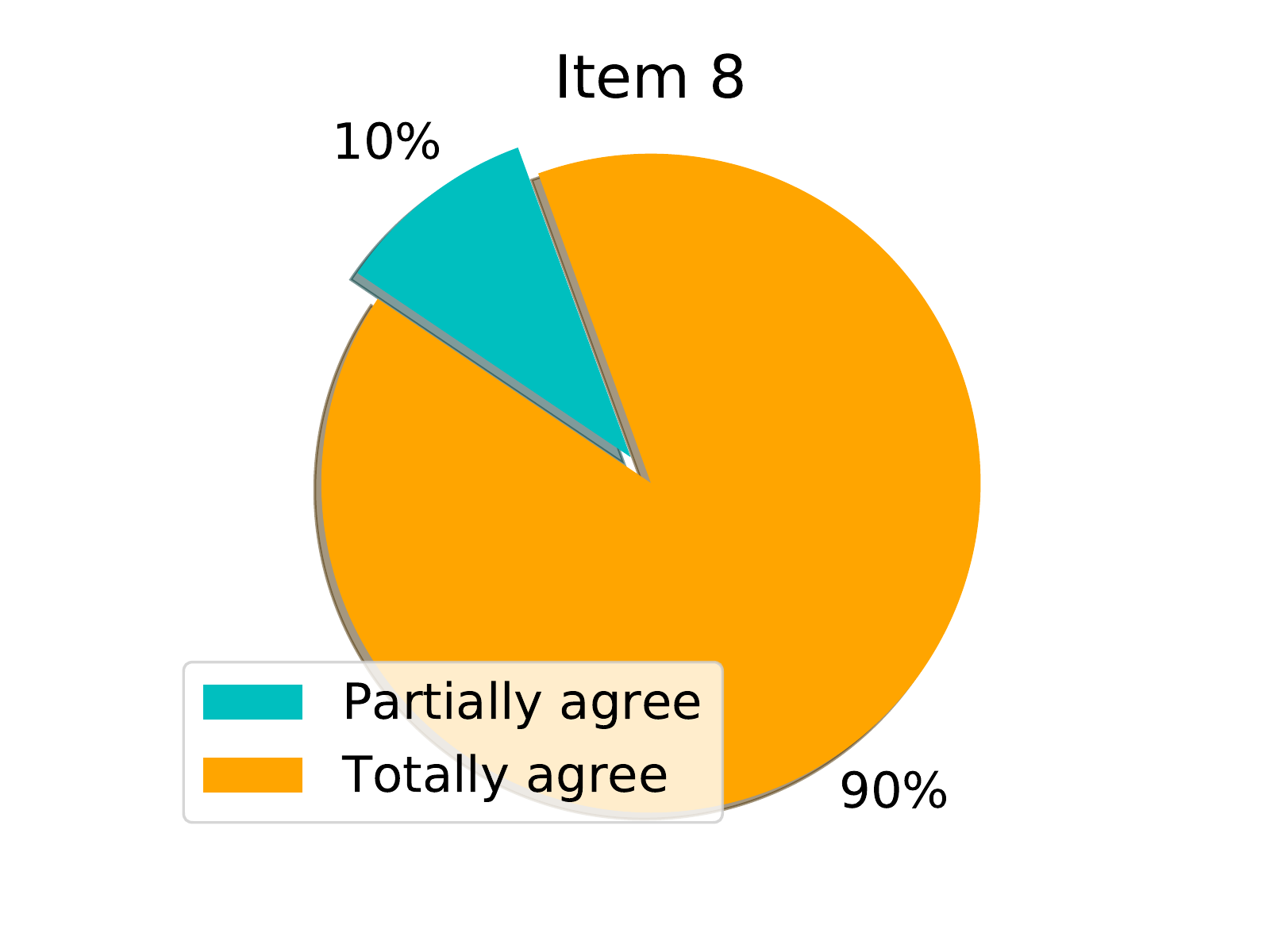}\includegraphics[scale=0.4]{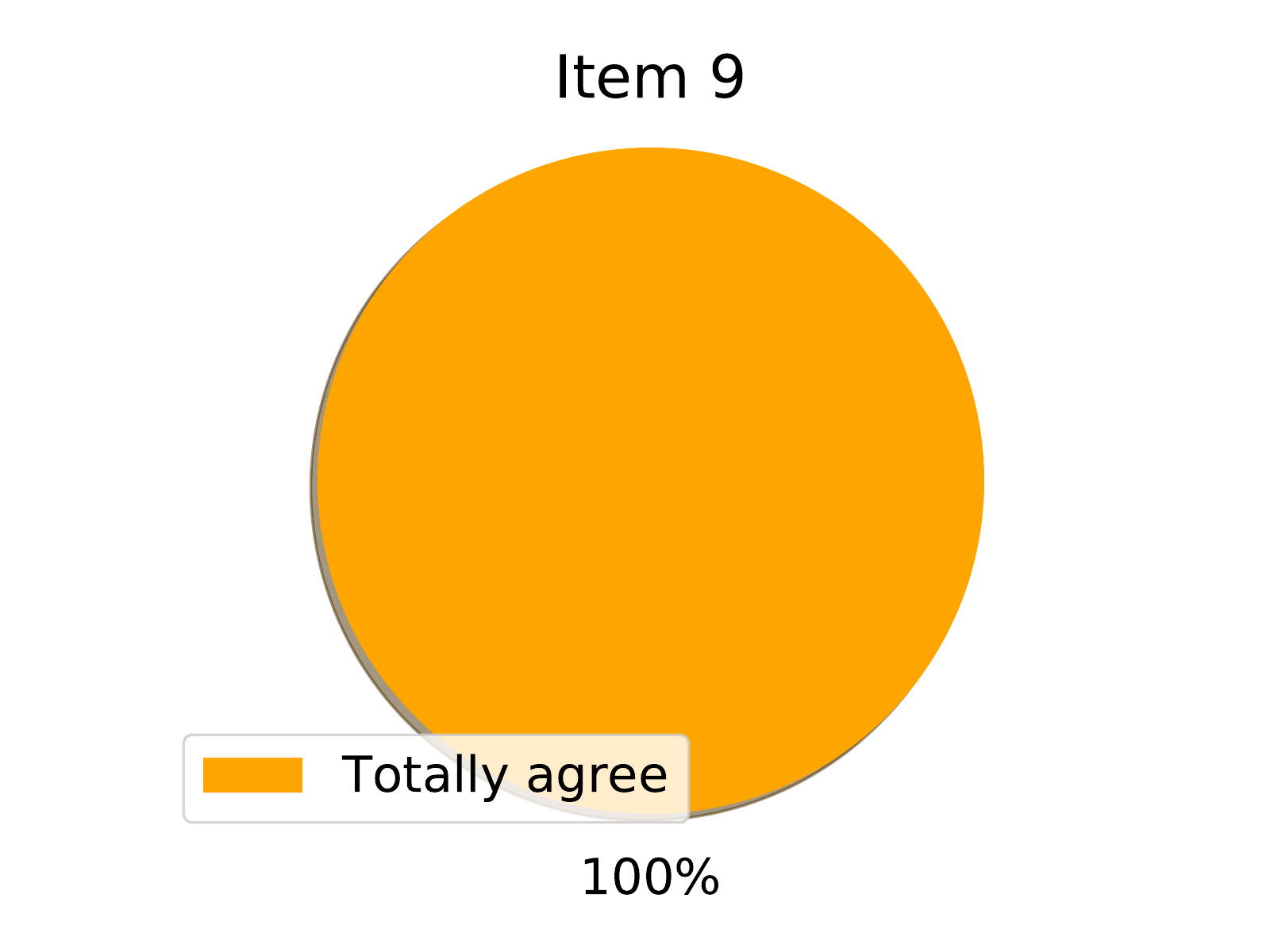}\\
	\includegraphics[scale=0.4]{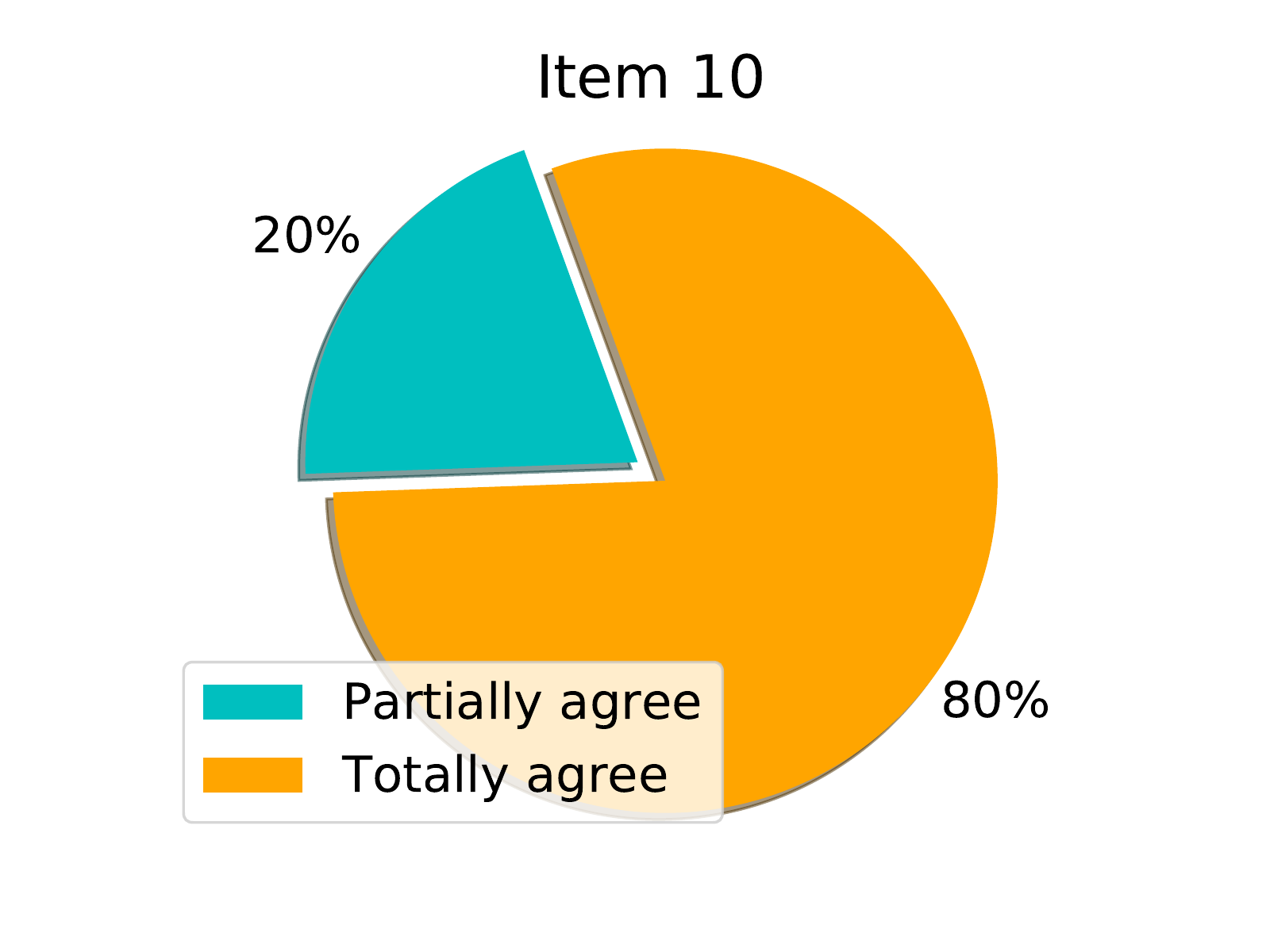}\includegraphics[scale=0.4]{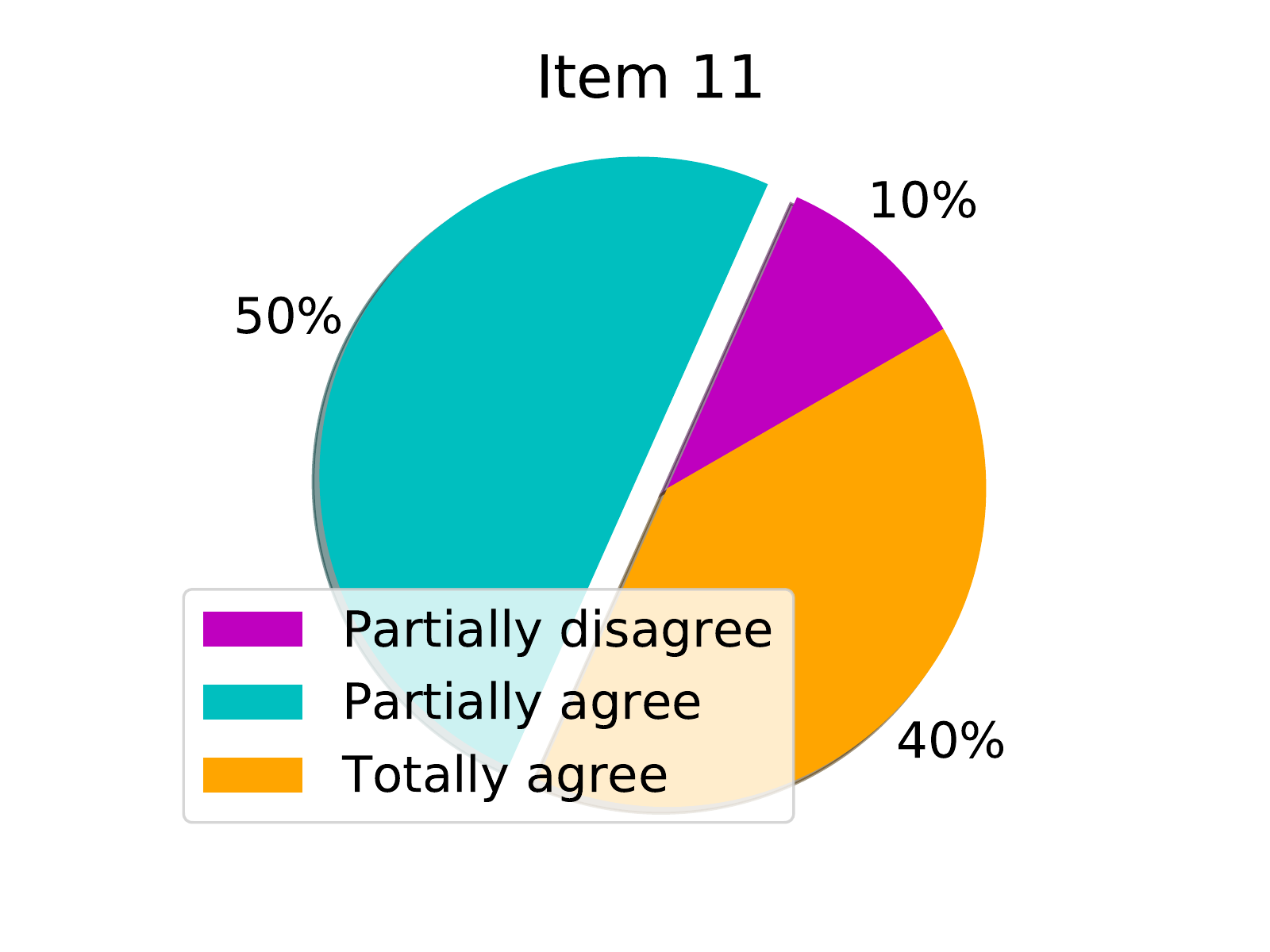}
	\caption{\label{fig3} \textbf{2\textsuperscript{st} Block of results}. The numbering of each of the graphs in this figure corresponds to the numbering of each question, from 7 to 11, of the applied questionnaire - see Appendix~\ref{Anx:A}.}
\end{figure*}

    The survey was submitted to a group of students in the discipline of Classical Mechanics (Physics vol.1), at the end of the set of steps described in the methodology - section \ref{sec:met} - and returned by them anonymously. In order to minimize the social effect of students answering questions fearful of a reevaluation of their grades, a second group of students received the survey in the semester following the mentioned discipline. The percentage results obtained from the students' responses to each item / statement in the survey are shown below. To divide into two blocks, the results for items 1 to 6 are grouped in figure \ref{fig2}, while those about items 7 to 11 in figure \ref{fig3}.

\subsection*{I.~~~1\textsuperscript{st}  Block of results - Items 1 to 6}
    We started our survey with a statement about the application of a new methodology, in order to know their perceptions that an active methodology would facilitate (or not) the fixation of the content seen in the classroom. The statement was placed as follows: {\em ``The proposed methodology allowed for a greater fixation of the theoretical content presented in the classroom''}. The chart {\bf Item 1}, in figure 2, shows the result, with 40\% of the students answering that they agree partially, while 60\% of them agree totally. 

	The concept explored in the second statement was the creativity. This ability is characterized by the ability to create, invent, innovate, both in the artistic and scientific fields. {\em ``As for the proposition that the student presents a project to verify the theory studied, this was important for each one to explore their creativity''}. The graph for {\bf Item 2} shows the result for this statement, in which 20\% of interviewees replied they partially agree, while 80\%  totally agree.

    In the third statement, {\em ''The proposed methodology stimulates the student's role as a protagonist in the face of the usual methodologies that use the application of a pre-determined script''}, we try to infer the students' perception of the main idea of an active methodology of taking on the student as a protagonist. As a result, shown in the chart {\bf Item 3}, 20\%  partially agree, while 80\%  totally agree.

    In the fourth statement, we approach the topic of critical sense: the ability to question and analyze in a rational and intelligent way. With this motivation (knowing the student if his critical sense was stimulated), the statement was: {\em ``The fact that there was a second seminar, after the experiment was carried out, stimulated the critical sense of the group, in order to argue, in a scientific way , the obtained results''}. As a result, we have the {\bf Item 4} graph, in which 10\% answered that they are indifferent to the questioning, 20\% partially agree and 70\% totally agree. 

    Knowing how to present and defend an idea (a position) is a skill of great value in the job market. The fifth item in the survey, {\em ``The methodology used is closer to the challenges you will encounter in the job market in terms of proposing and defending a project''}, approaches this concept, in order to know if the methodology could help the student in his preparation for the leadership of projects in the labor market. The {\bf Item 5} graph shows that only 10\% responded to being indifferent to the questioning, 20\% of them partially agree and 70\% agree completely.

    In carrying out the project, the student had the freedom to choose the theme of the work, as well as how and when to do it. What we wanted to evaluate in the next statement, {\em ``The fact that there is a high degree of freedom in carrying out the project can facilitate failures, such as a lack of responsibility''}, was whether the student was able to associate this high degree of freedom with the responsibility of executing the project or whether freedom was, at a certain point, a negative factor for the development of the activity. In the {\bf Item 6} graph, we see that 10\% of the interviewees responded that they totally disagree with the questioning, 40\% that partially disagree, 30\% partially agree and only 20\% totally agreed that excess of freedom was a bad factor, contributing to failures in the project.

\subsection*{II.~~~2\textsuperscript{nd}  Block of results - Items 7 to 11}
    Proactivity is one of the most important skills for the success of the methodology, given the oneself concept of active methodologies. Thus, we present the following statement to students: {\em ``The proposed methodology requires that all members of the group have a high degree of proactivity''}. That is, all members of the group must be fully participating in the proposed project. This type of statement contains a subjective charge, suggesting at least two interpretations for the responses obtained - see graph {\bf Item 7}. The first one takes into account that everyone in the group was proactive, so they agreed with the statement. The second interpretation would be taking into account that the lack of this proactivity may have occurred and that is why the students agreed with the statement. The result obtained was: 40\% answered that they partially agree and the other 60\% that totally agree. 

    The active methodology allows the student choose which path to follow in carrying out the activity, providing a stimulus to his critical sense and creativity. It is important to know if the student is motivating when he is treated as a protagonist of his own learning. In this sense, the following statement was made: {\em ``The way of carrying out the activities was motivating''}. The result obtained, shown in the graph {\bf Item 8}, shows 90\% of the students answering that they totally agree, while only 10\% answering that they partially agree. Note that this positive result includes students from the two groups interviewed.

    A question about time management was asked to students through the ninth question, which makes the following statement {\em ``Knowing how to manage time and divide tasks well is fundamental for the success of the project''}. The {\bf Item 9} graph shows that the result obtained was that 100\% of the students responded that they totally agree. This suggests how important it is to have the skill of time management and division of tasks, a skill that is widely required in the job market where the professional works in the model of goals and delivery of results. 

    The graph {\bf Item 10} refers to the statement: {\em ``Greater student interaction in the process of building one's own knowledge is the main characteristic of an approach using active teaching methodologies. The student starts to have more control and effective participation in the classroom, since it requires varied mental actions and constructions''}. This question aimed to verify whether the student understood that the methodology used during classes was an active methodology. The result was satisfactory, showing that the students were able to understand the purpose of the developed methodology: 80\% totally agree, while the others partially agree.

    A question arose during the writing of this article: Can this method be used in other disciplines? Disciplines that contain some risk to the student's physical integrity, for example, a practice involving an electrical circuit, in which the student will be in contact with sensitive items or that contain some eminent risk. Looking for answers to our questioning, we took the matter to the students through the following statement: {\em ``This type of methodology could be easily adopted in other disciplines''}. The result ({\bf Item 11}) obtained was that 10\% responded that they partially disagree, 50\% that partially agree and 40\% that totally agree. 

	The survey also included a space left free for students to express themselves in a written way. We leave here as an example, anonymously, the statement of one of the students who participated in this methodology:

\begin{quote}
    ``{\em When the seminar was proposed, although I was a little anxious, as it was the first academic work I was going to present at graduation, I thought it was a good idea and stopping to reflect after completing what the job was and what  it added to  me, I can say that it was of great value to me.

    As the seminar had the purpose of developing a work from scratch to prove a certain topic, it helped me to have a broader view on the topics covered and how I could put into practice the theoretical part that it is passed in class, instead of just following the steps of the experiments offered in the physics lab.
        
    There are countless ways for you to be able to demonstrate a certain theme, from the simplest to the most complex, and in the middle of this way of demonstrating you end up also deepening your theoretical basis, learning things you didn't know.
        
    As for the presentation, I was very tense in the first and a little less in the second presentation, I think this is a point where I need to improve and with the reservations [of the professor and laboratory technician], just as I think I was better in the second presentation than in the first, I intend to be better in the next one they propose to me.}''
\end{quote}

\section{Conclusions}
    In this work, we proposed a methodology to be applied in Physics laboratories (applied to engineering, in the case studied), as an alternative to the commonly used protocol: ({\em i}) theoretical exposition, ({\em ii}) exposure of the experimental script, ({\em iii}) execution of the experiment and ({\em iv}) report of the experimental activity. The guide for proposing this methodology was to encourage the student to be the protagonist in experimental activities, changing the logic and sequencing of these steps. Namely, we propose: ({\em i}) theoretical exposition, ({\em ii}) theoretical seminar and proposition of the experimental script and ({\em iii}) seminar for the exposition of the experiment carried out. Each of these steps is proposed with one or more professional skills as a guide, those frequently sought by companies and worked at jr Enterprises. and Startups, such as innovation, creativity, proactivity, protagonism, critical sense and scientific thinking.

    We applied our methodology to experimental physics classes and then we sought to learn from students  their positions regarding this dynamic proposed through a questionnaire, whose objective was to estimate whether they were able to observe and develop the skills that the method aims to stimulate, in order to approximate the professional's academic environment. In this sense, the method was able to make students think about the essential skills for an active methodology, even if they did not have prior knowledge about it.

    It is worth highlighting the connection between the proposed model diagram and the bases of the scientific method. If we take the fundamental steps of the scientific method, we can write that it essentially takes the steps: ({\em i}) Observation, ({\em ii}) Question, ({\em ii}) Research, ({\em iv}) Hypothesis, ({\em v}) Experiment, ({\em vi}) Analysis, ({\em vii}) Conclusion. In connection with our methodology, we can  associate these steps with the diagram shown in the figure \ref{FIG1} in which we have the stages of Theoretical Exposure associated with ({\em i}), Choice of theme linked to ({\em ii}), Experimental design to steps ({\em iii} - {\em v}) and Argumentative Exposure to ({\em vi}) and ({\em vii}). Unsurprisingly, the connection with the scientific method is readily a guide to proposals for active methodologies. 
    
    Our analysis can be extended to other disciplines with the reservations that the teacher attests to the safety of the nature of the experiments and that the proposed activities can be fully carried out by the students, in order to stimulate the protagonism of the students in their own learning process in the association between theory and practice.

\section{Acknowledgment}
    Authors would like to thanks ICEA/UFOP. RSF would like to thank the Aux{\'i}lio Pesquisador program/PROPP/UFOP and the National Council for Scientific and Technological Development - CNPq, under the process 424950/2018-9.

\appendix
\section{The Survey}
\label{Anx:A}
    For each item from 1 to 11 listed below, the interviewee must complete the parenthesis, according to his judgment, with the letters: (a) {\em Strongly disagree}; (b) {\em Partially disagree}; (c) {\em Indifferent}; (d) {\em Partially agree}; (e)  {\em Totally agree}.

\begin{enumerate}
    \item (\hspace{0.5cm}) The proposed methodology allowed a greater fixation of the theoretical content presented in the classroom.
 
    \item (\hspace{0.5cm}) As for the proposition that the student presents a project to verify the studied theory, this was important for each one to explore their creativity.
  
    \item (\hspace{0.5cm}) The proposed methodology stimulates the student's role as a protagonist compared to the usual methodologies that use the application of a predetermined script.
 
    \item (\hspace{0.5cm}) The fact that there was a second seminar, after the experiment was carried out, stimulated the critical sense of the group, in order to argue, in a scientific way, the results obtained.

    \item (\hspace{0.5cm}) The methodology used is closer to the challenges you will encounter in the job market in terms of proposing and defending a project.
 
    \item (\hspace{0.5cm}) The fact that there is a high degree of freedom in carrying out the project can facilitate failures, such as a lack of responsibility.
 
    \item (\hspace{0.5cm}) The proposed methodology requires that all members of the group have a high degree of proactivity.
 
    \item (\hspace{0.5cm}) The way of carrying out the activities was motivating.

    \item (\hspace{0.5cm}) Knowing how to manage time and divide tasks well is fundamental to the success of the project.
 
    \item (\hspace{0.5cm}) Greater student interaction in the process of building one's own knowledge is the main characteristic of an approach by active teaching methodologies. The student starts to have more control and effective participation in the classroom, since it requires varied mental actions and constructions.
 
    \item (\hspace{0.5cm}) This type of methodology could be easily adopted in other disciplines.
 
 \end{enumerate}

\end{multicols}
\end{document}